\documentclass[aps,pre,
twocolumn
]{revtex4-1}

\usepackage{mathrsfs}
\usepackage{bm}
\usepackage{amsmath}
\usepackage{amssymb}
\usepackage{graphicx}
\usepackage{esint}
\usepackage{mathrsfs}
\usepackage{bbm}

\makeatletter

\usepackage{hyperref}
\usepackage{color}
\definecolor{myblue}{RGB}{0,50,200}
\hypersetup{
colorlinks,
citecolor=myblue,
linkcolor=myblue,
urlcolor=myblue
}

\allowdisplaybreaks

\newcommand{\bms}{\mathbbmss}
\newcommand{\mca}{\mathcal}
\newcommand{\mbb}{\mathbb}
\newcommand{\mrm}{\mathrm}
\newcommand{\avg}[1]{\langle #1\rangle}

\newcommand{\avgl}[1]{\left\langle #1\right\rangle}

\newcommand{\bra}[1]{\left( #1 \right)}
\newcommand{\bras}[1]{\left[ #1 \right]}
\newcommand{\brab}[1]{\left\{ #1 \right\}}

\newcommand{\pp}{\partial}
\newcommand{\br}{\bm{r}}
\newcommand{\bv}{\bm{v}}
\newcommand{\xs}{\bm{r},\bm{v}}
\newcommand{\dxs}{d\bm{r}d\bm{v}}
\newcommand{\mxs}{\bm{r},\bm{v}/(1+\theta)}
\newcommand{\pss}{P^{\rm ss}}
\newcommand{\mpss}{P_{\theta}^{\rm ss}}
\newcommand{\bL}{\bm{\Lambda}}
\newcommand{\bG}{\bm{\Gamma}}
\newcommand{\bram}[1]{\bigg( #1 \bigg)}
\newcommand{\brah}[1]{\Bigg( #1 \Bigg)}
\newcommand{\bT}{\bm{\Theta}}

\begin{document}
\title{Uncertainty relations for underdamped Langevin dynamics}

\author{Tan Van Vu}
\email{tan@biom.t.u-tokyo.ac.jp}

\affiliation{Department of Information and Communication Engineering, Graduate
School of Information Science and Technology, The University of Tokyo,
Tokyo 113-8656, Japan}

\author{Yoshihiko Hasegawa}
\email{hasegawa@biom.t.u-tokyo.ac.jp}

\affiliation{Department of Information and Communication Engineering, Graduate
School of Information Science and Technology, The University of Tokyo,
Tokyo 113-8656, Japan}

\date{\today}

\begin{abstract}
A trade-off between the precision of an arbitrary current and the dissipation, known as the thermodynamic uncertainty relation, has been investigated for various Markovian systems.
Here, we study the thermodynamic uncertainty relation for underdamped Langevin dynamics.
By employing information inequalities, we prove that, for such systems, the relative fluctuation of a current at a steady state is constrained by both the entropy production and the average dynamical activity.
We find that, unlike what is the case for overdamped dynamics, the dynamical activity plays an important role in the bound.
We illustrate our results with two systems, a single-well potential system and a periodically driven Brownian particle model, and numerically verify the inequalities.
\end{abstract}

\maketitle

\section{Introduction}

Fluctuations in small systems far from thermal equilibrium have been actively studied over the last two decades.
Universal relations that characterize nonequilibrium systems, such as fluctuation-dissipation theorems \cite{Kubo.2012,Paolo.2007.PRL,Prost.2009.PRL,Lippiello.2014.PRL} and fluctuation theorems \cite{Gallavotti.1995.PRL,Jarzynski.1997.PRL,Seifert.2012.RPP}, have been discovered as a result.
These relations are fundamentally important and enable us to investigate the properties of many physical systems experimentally.
In recent years, the thermodynamic uncertainty relation (TUR) \cite{Barato.2015.PRL,Gingrich.2016.PRL}, quantifying a trade-off between the fluctuation of an arbitrary current and the dissipation, has been an important discovery in statistical physics.
Qualitatively, the TUR indicates that it is impossible to attain a small fluctuation without increasing the dissipation quantified by entropy production in the system.
In particular, it states that
\begin{equation}
\frac{\mrm{Var}[\Theta]}{\avg{\Theta}^2}\ge \frac{2}{\mca{T}\sigma}\label{eq:conv.TUR}
\end{equation}
holds for an arbitrary current $\Theta$ and finite observation time $\mca{T}$ in a nonequilibrium steady state.
Here, $\avg{\Theta}$ and $\mrm{Var}[\Theta]$ denote the mean and variance of the current, respectively, and $\sigma$ is the entropy production rate.
Hereafter, the term ``bound'' refers to the lower bound on the current fluctuation and $2/(\mca{T}\sigma)$ is called the original bound.

The TUR has been proven for continuous-time discrete-state Markov-jump processes \cite{Horowitz.2017.PRE}, extended to other contexts \cite{Pietzonka.2016.PRE,Polettini.2016.PRE,Garrahan.2017.PRE,Proesmans.2017.EPL,Hyeon.2017.PRE,Rotskoff.2017.PRE,Barato.2018.NJP,Barato.2019.JSM,Macieszczak.2018.PRL,Brandner.2018.PRL,Hasegawa.2019.PRE,Koyuk.2019.JPA,Vu.2019.PRE}, and successfully applied to study a range of specific problems \cite{Barato.2015.JPCB,Falasco.2016.PRE,Patrick.2016.JSM,Gingrich.2017.PRL,Hwang.2018.JPCL}.
A remarkable application of the TUR is in the quantification of the entropy production from experimental data using fluctuating currents \cite{Li.2019.NC}.
It is of particular interest that the TUR is also valid for overdamped Langevin systems \cite{Pigolotti.2017.PRL,Dechant.2018.JSM}.
However, overdamped dynamics, which are only approximate descriptions of underdamped dynamics, can dramatically fail to capture thermodynamic quantities such as the entropy production \cite{Matsuo.2000.PA,Celani.2012.PRL,Bo.2014.JSP}.
Therefore, it is natural to ask whether the TUR is valid for underdamped systems; and more importantly, if not, whether there exists an analogous bound for such systems.

In the present paper, we address the aforementioned question for underdamped Langevin dynamics, wherein damping does not suppress inertial effects.
Regarding the validity of the TUR, we numerically found that it does not universally hold.
Therefore, the original bound cannot be utilized for general underdamped systems.
By applying information inequalities to systems, we derive new bounds for both scalar and vector currents in the steady state.
Specifically, we prove that the relative fluctuation of a current is bounded from below by a quantity involving entropy production and the average dynamical activity.
This activity term is a kinetic aspect of the system and plays a central role in characterizing its dynamics \cite{Lecomte.2005.PRL,Merolle.2005.PNAS,Lecomte.2007.JSP,Garrahan.2009.JPA,Maes.2017.PRL,Shiraishi.2018.PRL}.
The results imply that the current fluctuation is constrained not only by entropy production but also by dynamical activity.
We empirically verify the derived bounds via numerical simulations.
In addition, we show that our approach can be applied to the derivation of an uncertainty relation for active matter systems, which have recently attracted considerable interest \cite{Fily.2012.PRL,Marchetti.2013.RMP,Speck.2014.PRL,Farage.2015.PRE,Marini.2015.SM,Fodor.2016.PRL,Mandal.2017.PRL,Nardini.2017.PRX}.

\section{Model}

We consider a general underdamped system of $N$ particles, wherein the particle $i$ is in contact with a heat reservoir in equilibrium at temperature $T_i$.
The dynamics of the system are described by a set of coupled equations as follows:
\begin{equation}
\dot{r}_i=v_i,~m_i\dot{v}_i=-\gamma_iv_i+F_i(\br)+\xi_i,\label{eq:gen.und.Lan}
\end{equation}
where the dots indicate time derivatives, $\br=[r_1,\dots,r_N]^\top$ and $\bv=[v_1,\dots,v_N]^\top$ denote positions and velocities, respectively; $m_i$ and $\gamma_i$ are the mass and the damping coefficient of particle $i$, respectively; $F_{i}(\br)=-\pp_{r_i}U(\br)+f_i(\br)$ is the total acting force with the potential $U(\br)$; and the $\xi_i$'s are zero-mean white Gaussian noises with variances $\avg{{\xi}_{i}(t){\xi}_{j}(t')}=2T_i\gamma_i\delta_{ij}\delta(t-t')$.
Throughout this study, Boltzmann's constant is set to $k_{\rm B}=1$.
The time evolution of the phase-space probability distribution function, $P(\xs,t)$, can be described by the Kramers equation
\begin{equation}
\pp_tP(\xs,t)=\sum_{i=1}^N\bras{-\pp_{r_i}J_{r_i}(\xs,t)-\pp_{v_i}J_{v_i}(\xs,t)},\label{eq:Kramers.equ}
\end{equation}
where $J_{r_i}(\xs,t)=v_iP(\xs,t)$ and $J_{v_i}(\xs,t)=1/m_i\bras{-\gamma_iv_i+F_i(\br)-T_i\gamma_i/m_i\pp_{v_i}}P(\xs,t)$.
Hereafter, we focus exclusively upon the steady state, in which the system has stationary distribution $\pss(\xs)$ and probability currents $J^{\rm ss}_{r_i}(\xs)$ and $J^{\rm ss}_{v_i}(\xs)$.

Let $\bm{\Gamma}\equiv\bras{\br(t),\bv(t)}_{t=0}^{t=\mca{T}}$ denote a phase-space trajectory that starts at the point $\bra{\br^0,\bv^0}\equiv\bra{\br(0),\bv(0)}$ and has length $\mca{T}$.
The entropy production characterizes the irreversibility in the system and has been generalized to the level of stochastic trajectories \cite{Seifert.2005.PRL}.
Along the trajectory $\bm{\Gamma}$, the entropy production can be defined as $\Delta s_{\rm tot}[\bm{\Gamma}]\equiv\ln\bra{\bms{P}[\bm{\Gamma}]/\bms{P}^\dagger[\bm{\Gamma}^{\dagger}]}$,
which is a comparison of the probabilities of the forward path $\bm{\Gamma}$ and its time-reversed counterpart $\bm{\Gamma}^{\dagger}\equiv\bras{\br(\mca{T}-t),-\bv(\mca{T}-t)}_{t=0}^{t=\mca{T}}$ \cite{Lee.2013.PRL,Spinney.2012.PRE,Spinney.2012.PRL}.
Here, $\bms{P}[\bm{\Gamma}]$ and $\bms{P}^\dagger[\bm{\Gamma}^{\dagger}]$ are the probabilities of observing the forward path $\bm{\Gamma}$ and its time-reversed path $\bm{\Gamma}^{\dagger}$, respectively.
Because the initial distribution in the time-reversed process is $P(\br(\mca{T}),\bv(\mca{T}),\mca{T})$, the entropy production can be decomposed as
\begin{equation}\label{eq:ent.decomp}
\Delta s_{\rm tot}[\bm{\Gamma}]=-\ln\frac{P(\br(\mca{T}),\bv(\mca{T}),\mca{T})}{P(\br(0),\bv(0),0)}+\ln\frac{\bms{P}[\bm{\Gamma}|\br(0),\bv(0)]}{\bms{P}[\bm{\Gamma}^{\dagger}|\br(\mca{T}),-\bv(\mca{T})]},
\end{equation}
where the first and second terms in the right-hand side of Eq.~\eqref{eq:ent.decomp} correspond to the system entropy production and the medium entropy production, respectively.
Using stochastic thermodynamics \cite{Seifert.2012.RPP,Spinney.2012.PRL}, we can show that the entropy production rate, which includes changes in the system entropy and the medium entropy, is as follows:
\begin{equation}
\sigma=\sum_{i=1}^{N}\iint\dxs\frac{m_i^2}{T_i\gamma_i}\frac{J_{v_i}^{\rm ir}(\xs)^2}{\pss(\xs)},\label{eq:ent.form}
\end{equation}
where $J_{v_i}^{\rm ir}(\xs)$ is the irreversible current given by $J_{v_i}^{\rm ir}(\xs)=-1/m_i\bras{\gamma_iv_i+T_i\gamma_i/m_i\pp_{v_i}}\pss(\xs)$.
The detailed calculation of $\sigma$ is provided in Appendix \ref{app:ent.cal}.
For an arbitrary trajectory $\bm{\Gamma}$, we consider a generalized observable-type current, $\Theta[\bm{\Gamma}]=\int_0^{\mca{T}}dt\,\bm{\Lambda}(\br)^\top\circ\dot{\br}$, where $\bm{\Lambda}\in\mbb{R}^{N\times 1}$ is the projection function and $\circ$ denotes the Stratonovich product.
Our aim is to derive a lower bound on the fluctuation of the current $\Theta[\bm{\Gamma}]$.

\section{Uncertainty relations}
\subsection{Derivation}
To derive our results, let us consider the auxiliary dynamics
\begin{equation}
\dot{r}_i=v_i,~m_i\dot{v}_i=H_{i,\theta}(\br,\bv)+\xi_i,\label{eq:aux.und.Lan}
\end{equation}
where $\theta$ is a perturbation parameter and $H_{i,\theta}(\br,\bv)$ is the force acting upon particle $i$.
The detailed form of $H_{i,\theta}(\br,\bv)$ will be determined later.
For an arbitrary function $f[\bm{\Gamma}]$, let $\avg{f}_{\theta}\equiv\int\mca{D}\bm{\Gamma}\,f[\bm{\Gamma}]\bms{P}_{\theta}[\bm{\Gamma}]$ and $\mrm{Var}_\theta[f]\equiv\avg{\bra{f-\avg{f}_\theta}^2}_\theta$.
Here, $\bms{P}_{\theta}[\bm{\Gamma}]$ is the probability of observing the trajectory $\bm{\Gamma}$ generated by the auxiliary dynamics.
It can be expressed in a path-integral form as \cite{Imparato.2006.PRE}
\begin{equation}
\bms{P}_\theta[\bm{\Gamma}]=\mca{N}\pss_\theta(\br^0,\bv^0)\prod_{i=1}^{N}\exp\bra{-\mca{A}_i[\bm{\Gamma}]},\label{eq:pat.int}
\end{equation}
where $\mca{A}_i[\bm{\Gamma}]\equiv\int_{0}^{\mca{T}}dt\bra{m_i\dot{v}_i-H_{i,\theta}(\br,\bv)}^2/(4T_i\gamma_i)$ is an Onsager--Machlup action functional, $\pss_\theta(\br^0,\bv^0)$ is the stationary distribution of the auxiliary dynamics, and $\mca{N}$ is a term independent of $\theta$.
The integral in the action functional is interpreted as the continuum limit of an Ito sum with pre-point discretization.
Note that writing the crossing term $\int dt\,\dot{v}_iH_{\theta,i}(\br,\bv)$ in Stratonovich integral (i.e., mid-point discretization) results in a different form of the path integral.
However, it can be shown that both pre- and mid-point discretization schemes reduce to the same path-integral representation in the case of additive noise (see Appendix \ref{app:pat.int}).
Hereafter, the notations $\avg{\cdot\cdot}$ and $\avg{\cdot\cdot}_\theta$ imply averages taken over ensembles in the original and auxiliary dynamics, respectively.
In the steady state, the average of the current becomes $\avg{\Theta}_\theta=\mca{T}\iint\dxs\,\bm{\Lambda}(\br)^\top \bm{J}_{r,\theta}^{\rm ss}(\xs)$, where $\bm{J}_{r,\theta}^{\rm ss}(\xs)=\bv\pss_\theta(\xs)$ is the vector of probability currents in the auxiliary dynamics.
Since $\avg{\Theta}_{\theta}$ is a function of $\theta$, i.e., $\avg{\Theta}_{\theta}=\psi(\theta)$, we can consider $\Theta$ as an estimator of $\psi(\theta)$.
The precision of this estimator is bounded from below by the reciprocal of the Fisher information as \cite{Hasegawa.2019.PRE}
\begin{equation}
\frac{\mrm{Var}_{\theta}\bras{\Theta}}{\bra{\pp_\theta\avg{\Theta}_\theta}^2}\ge\frac{1}{\mca{I}(\theta)},\label{eq:TCRI}
\end{equation}
where $\mca{I}(\theta)\equiv-\avgl{\pp_\theta^2\ln\bms{P}_\theta}_\theta$ is the Fisher information, which can be calculated via the path integral in Eq.~\eqref{eq:pat.int}.
Equation~\eqref{eq:TCRI} is known as the Cram{\'e}r--Rao inequality, and indicates a trade-off between the precision of an estimator and the Fisher information.
The analogy between this inequality and the TUR has been utilized to derive the original bound \cite{Hasegawa.2019.PRE,Dechant.2019.JPA}.
Next, we use the virtual-perturbation technique \cite{Dechant.2018.arxiv} and consider the auxiliary dynamics with the following force:
\begin{equation}
\label{eq:aux.drift}
\begin{aligned}
H_{i,\theta}&(\xs)=-(1+\theta)\gamma_iv_i+(1+\theta)^2F_i(\br)\\
&+\frac{T_i\gamma_i}{m_i}(1-(1+\theta)^3)\frac{\pp_{v_i}\pss(\mxs)}{\pss(\mxs)}.
\end{aligned}
\end{equation}
When $\theta=0$, $H_{i,\theta}(\xs)=-\gamma_iv_i+F_i(\br)$ and the auxiliary dynamics become the original ones.
Verifying that these auxiliary dynamics have a stationary distribution $P_\theta^{\rm ss}(\xs)=P^{\rm ss}(\mxs)/(1+\theta)^N$ is easy.
The average current in the auxiliary dynamics is related to that in the original dynamics as $\avg{\Theta}_\theta=(1+\theta)\avg{\Theta}$; thus, we have $\pp_\theta\avg{\Theta}_\theta=\avg{\Theta}$.
By letting $\theta=0$ in Eq.~\eqref{eq:TCRI}, we obtain the following inequality:
\begin{equation}
\frac{\mrm{Var}[\Theta]}{\avg{\Theta}^2}\geq\frac{2}{\Sigma},\label{eq:gen.TUR}
\end{equation}
where $\Sigma=2\mca{I}(0)=\mca{T}\bra{9\sigma+4\Upsilon}+\Omega$ and
\begin{align}
\Upsilon&=\sum_{i=1}^N\Big(\frac{1}{T_i\gamma_i}\avg{F_i(\br)^2}-3\frac{\gamma_i}{T_i}\avg{v_i^2}+4\frac{\gamma_i}{m_i}\Big),\\
\Omega&=2\bigg\langle{\Big(\sum_{i=1}^{N}v_i\pp_{v_i}\pss(\xs)/\pss(\xs)\Big)^2}\bigg\rangle-2N^2.
\end{align}
Inequality \eqref{eq:gen.TUR} is the main result of our paper and holds for an arbitrary time scale and general underdamped systems.
The detailed derivation is provided in Appendix \ref{app:TUR.deriv}.
As seen, the derived bound is not equal to the reciprocal of entropy production as is the original bound (which will be shown to be violated in underdamped systems).
In addition to the entropy production, the derived bound also contains $\Upsilon$, which involves the moments of the forces and velocities, and a boundary term $\Omega$, which is always nonnegative, as $\Omega=2\avg{\bras{\pp_\theta\ln P_\theta^{\rm ss}(\br^0,\bv^0)}^2}_{\theta=0}$.
In the long-time limit, i.e., $\mca{T}\to\infty$, the boundary term can be neglected. Thus, our result reduces to a bound that was derived in Ref.~\cite{Fischer.2018.PRE} for a one-dimensional system using large deviation theory.
Since the kinetic term $\Upsilon$ can be estimated from the experimental data, our bound can be applied to quantify the entropy production for underdamped systems as in Ref.~\cite{Li.2019.NC}, where the original bound has been used to infer a lower bound on entropy production for overdamped systems.

Considering the equality condition of the derived bound, the lower bound in Eq.~\eqref{eq:gen.TUR} can be attained if and only if the equality condition in Eq.~\eqref{eq:TCRI} is satisfied with $\theta=0$.
Equivalently, $\left.\pp_\theta\ln\bms{P}_\theta[\bm{\Gamma}]\right|_{\theta=0}=\mu\bra{\Theta[\bm{\Gamma}]-\psi(0)}$ must hold for an arbitrary trajectory $\bm{\Gamma}$, where $\mu$ is a scaling coefficient.
However, it can be proven that this lower bound cannot be attained (the detailed proof is provided in Appendix \ref{app:equ.cond}).

So far, we have considered currents with projection functions involving only $\br$.
Here, we derive a bound for a certain class of more general currents, $\Theta[\bG]=\int_0^{\mca{T}}dt\,\bL(\xs)^\top\circ\dot{\br}$, wherein the projection function involves the velocity variables.
Suppose that $\bL(\xs)$ satisfies $\bL(\br,(1+\theta)\bv)=(1+\theta)^d\bL(\xs)$ for all $\theta\in\mbb{R}$ and for some nonnegative number $d$.
For example, $\bL(\xs)$ can be a vector of homogeneous polynomials of $\bv$ (i.e., all nonzero terms have the same degree $d$ with respect to $\bv$).
From the relationship $\avg{\Theta}_\theta=(1+\theta)^{d+1}\avg{\Theta}$, we have $\left.\pp_\theta\avg{\Theta}_\theta\right|_{\theta=0}=(d+1)\avg{\Theta}$.
Consequently, we obtain the following bound:
\begin{equation}
\frac{\mrm{Var}[\Theta]}{\avg{\Theta}^2}\ge\frac{2(d+1)^2}{\Sigma}.\label{eq:tig.TUR}
\end{equation}
When the projection function contains only $\br$, i.e., $d=0$, Eq.~\eqref{eq:tig.TUR} reduces to Eq.~\eqref{eq:gen.TUR}.

\subsection{Interpretation of the bound}

Now, we interpret the physical meaning of the term $\Upsilon$ in the bound.
In the original dynamics, the action functional in the path integral with mid-point discretization is \cite{Rosinberg.2016.EPL}
\begin{equation}
\mca{A}_i^{\rm m}[\bm{\Gamma}]\equiv\int_{0}^{\mca{T}}dt\bras{\frac{1}{4T_i\gamma_i}\bra{m_i\dot{v}_i+\gamma_iv_i-F_i}^2-\frac{\gamma_i}{2m_i}}.
\end{equation}
The functional $\mca{A}_i^{\rm m}$ can be decomposed into two contributions: the time-antisymmetric $(\mca{S})$ and time-symmetric $(\mca{K}_i,~\mca{K}_i^*)$ components as $-\mca{A}_i^{\rm m}[\bm{\Gamma}]=\mca{S}_i[\bm{\Gamma}]+\mca{K}_i[\bm{\Gamma}]+\mca{K}_i^*[\bm{\Gamma}]$ \cite{Falasco.2016.NJP}, where
\begin{align}
\mca{S}_i[\bm{\Gamma}]&=-\int_0^{\mca{T}}dt\,\frac{1}{2T_i}\bra{m_i\dot{v}_i-F_i}\circ v_i,\nonumber\\
\mca{K}_i[\bm{\Gamma}]&=\int_0^{\mca{T}}dt\bras{\frac{1}{4T_i\gamma_i}\bra{2m_i\dot{v}_i\circ F_i-\gamma_i^2v_i^2-F_i^2}+\frac{\gamma_i}{2m_i}},\nonumber\\
\mca{K}_i^*[\bm{\Gamma}]&=-\int_0^{\mca{T}}dt\,\frac{m_i^2\dot{v}_i^2}{4T_i\gamma_i}.
\end{align}
The time-antisymmetric part corresponds to the integrated entropy flux from the system into the reservoirs and is thermodynamically consistent with the definition of medium entropy production $\Delta s_{\rm m}[\bm{\Gamma}]\equiv\sum_{i=1}^N\frac{1}{T_i}\int_{0}^{\mca{T}}dt\,\bra{\gamma_i v_i-\xi_i}\circ v_i$.
The kinetic term $\mca{K}_i^*$ relates to the mean square acceleration of the particles and may quantify an amount of activity.
However, $\mca{K}_i^*$ should be interpreted as part of the functional measure; it determines the functional space over which to integrate \cite{Zinn.2002.CP}.
In particular, $\mca{K}_i^*$ collects trajectories for which $dv_i^2/dt$ remains finite when $dt\to 0$.
The time-symmetric term $\mca{K}_i$ is identified as the dynamical activity, which has been introduced in the literature \cite{Jack.2006.JCP,Roeck.2007.PRE,Baiesi.2009.JSP,Jack.2014.JPA,Maes.2014.NJP}.
In discrete-state Markov-jump processes, the dynamical activity characterizes the time scale of the system and serves as an essential term in the speed-limit inequality \cite{Shiraishi.2018.PRL} and the kinetic uncertainty relation \cite{Terlizzi.2019.JPA}.
Taking the average of $\mca{K}_i$, we explicitly obtain
\begin{equation}
\avg{\mca{K}_i}=\frac{\mca{T}}{4}\bra{\frac{1}{T_i\gamma_i}\avg{F_i(\br)^2}-3\frac{\gamma_i}{T_i}\avg{v_i^2}+4\frac{\gamma_i}{m_i}}.
\end{equation}
It can be easily confirmed that $\Upsilon=4\sum_{i=1}^N\avg{\mca{K}_i}/\mca{T}$. 
Therefore, the term $\Upsilon$ in the derived bound is exactly the average dynamical activity.
This implies that the current fluctuation in underdamped systems is not only constrained by the entropy production but also by its dynamical activity.

To clarify the role of dynamical activity in the bound, let us consider equilibrium systems (i.e., the external force $f_i(\br)=0$ and the temperature $T_i=T$ for all $i=1,\dots,N$), where the entropy production vanishes.
The steady-state distribution is of a Maxwell--Boltzmann type,
\begin{equation}
\pss(\xs)=\mca{C}\exp\bras{-\frac{1}{T}\bra{\frac{1}{2}\sum_{i=1}^Nm_iv_i^2+U(\br)}},
\end{equation}
where $\mca{C}$ is the normalizing constant.
The average dynamical activity is always positive, i.e., $\avg{\mca{K}_i}=\mca{T}\bra{\avg{F_i(\br)^2}/(T\gamma_i)+\gamma_i/m_i}/4>0$.
The average current does not always vanish as in the equilibrium overdamped systems, for example, for $\bL(\br,\bv)=\bv$.
Therefore, the fluctuation of a current in equilibrium is bounded only by the average dynamical activity.

In the following, we provide an explanation regarding why the dynamical activity appears in the lower bound of the current fluctuation in underdamped dynamics.
For general Langevin systems, the probability currents can be decomposed into irreversible and reversible components \cite{Spinney.2012.PRE}.
In overdamped dynamics, which involves only the even variables, reversible currents are absent from the decomposition and the probability currents are irreversible.
Both the entropy production and current fluctuation can be characterized via these probability currents.
Therefore, the current fluctuation can be bounded solely by the associated entropy production.
However, this is not the case for underdamped dynamics.
Unlike in the case of overdamped dynamics, both irreversible and reversible currents exist in the decomposition of underdamped dynamics.
Both types of currents jointly characterize the current fluctuation.
Moreover, the entropy production is quantified via the irreversible currents, as in Eq.~\eqref{eq:ent.form}, and the dynamical activity is quantified via the reversible currents.
Consequently, in underdamped dynamics, the current fluctuation is constrained not only by the entropy production but also by the dynamical activity.

\subsection{Multidimensional TUR}

Generally, there are correlations between currents in real-world systems.
Simultaneously observing multiple currents is expected to yield more statistical information concerning the distribution.
Therefore, the multidimensional TUR, which includes several currents in the observable, provides a tighter bound than does scalar TUR \cite{Dechant.2019.JPA}.
Such a bound can be applied to a study of the trade-off relationship between power and efficiency in steady-state heat engines \cite{Pietzonka.2018.PRL,Dechant.2019.JPA}.
Here, we derive the multidimensional TUR for underdamped systems.
We consider a vector observable $\bm{\Theta}\in\mbb{R}^{M\times 1}$, defined by
$\bm{\Theta}[\bm{\Gamma}]=\int_0^{\mca{T}}dt\,\bm{\Lambda}(\br)\circ\dot{\br}$, where $\bm{\Lambda}\in\mbb{R}^{M\times N}$ is an arbitrary matrix-valued function of $\br$.
By applying the information inequality for a multivariate estimator, we obtain (see Appendix \ref{app:eq.mul.CRI})
\begin{equation}\label{eq:mul.CRI}
\mrm{Cov}_{\theta}[\bm{\Theta}]\succeq\mca{I}(\theta)^{-1}\pp_{\theta}\avg{\bm{\Theta}}_{\theta}\pp_{\theta}\avg{\bm{\Theta}}_{\theta}^\top,
\end{equation}
where $\mrm{Cov}_{\theta}[\bm{\Theta}]=[\mrm{Cov}_{\theta}[\Theta_i;\Theta_j]]\in\mbb{R}^{M\times M}$ is the covariance matrix of the estimator $\bm{\Theta}$ and the matrix inequality $\bm{X}\succeq\bm{Y}$ indicates that $\bm{X}-\bm{Y}$ is positive semi-definite.
Here, $\mrm{Cov}_{\theta}[\Theta_i;\Theta_j]=\avg{\Theta_i\Theta_j}_\theta-\avg{\Theta_i}_\theta\avg{\Theta_j}_\theta$.
Considering the same auxiliary dynamics as in Eq.~\eqref{eq:aux.drift}, we obtain the multidimensional TUR as
\begin{equation}
\mrm{Cov}[\bm{\Theta}]\succeq\frac{2}{\Sigma}\avg{\bm{\Theta}}\avg{\bm{\Theta}}^\top.\label{eq:mul.TUR.raw}
\end{equation}
From Eq.~\eqref{eq:mul.TUR.raw}, one can derive various bounds for current fluctuation.
For instance, in the case of two-dimensional currents, i.e., $\bm{\Theta}=[\Theta_1,\Theta_2]^\top$, the condition of the positive semi-definite matrix yields a tighter bound:
\begin{equation}
\mrm{Var}[\Theta_1]\ge\frac{2}{\Sigma}\avg{\Theta_1}^2+\sup_{\Theta_2}\frac{\bra{\mrm{Cov}[\Theta_1;\Theta_2]-\frac{2}{\Sigma}\avg{\Theta_1}\avg{\Theta_2}}^2}{\mrm{Var}[\Theta_2]-\frac{2}{\Sigma}\avg{\Theta_2}^2}.\label{eq:two.obs.TUR}
\end{equation}
In addition to the variance of individual currents, this inequality also involves the correlation of two currents.
A remarkable point in Eq.~\eqref{eq:two.obs.TUR} is that, if a current $\Theta_1$ satisfies $\mrm{Var}[\Theta_1]=2\avg{\Theta_1}^2/\Sigma$, then $\mrm{Cov}[\Theta_1;\Theta_2]=2\avg{\Theta_1}\avg{\Theta_2}/\Sigma$ holds for an arbitrary nonvanishing current $\Theta_2$.
Although the equality condition of the derived bound cannot be attained, inequalities like Eq.~\eqref{eq:two.obs.TUR} provide insight into the correlation of currents.
In particular, for overdamped dynamics, the current $\Theta_{\rm tot}$ of stochastic total entropy production with a specific form of the drift function satisfies the equality condition \cite{Hasegawa.2019.PRE}; thus, $\mrm{Cov}[\Theta_{\rm tot};\Theta]=2\avg{\Theta}$ for an arbitrary current $\Theta$, which expresses a universal property of stochastic entropy production.
Another direction is to derive an inequality directly from the definition of the positive semi-definite matrix, $\bm{x}^\top\bra{\mrm{Cov}[\bm{\Theta}]-2\avg{\bm{\Theta}}\avg{\bm{\Theta}}^\top/\Sigma}^{-1}\bm{x}\geq 0$ for all $\bm{x}\in\mbb{R}^{M\times 1}$ (if $\bm{X}$ is a non-singular, positive semi-definite matrix, then so is $\bm{X}^{-1}$).
By choosing $\bm{x}=\avg{\bm{\Theta}}$, we obtain the following inequality (see Appendix \ref{app:eq.mul.TUR}):
\begin{equation}
\avg{\bm{\Theta}}^\top\mrm{Cov}[\bm{\Theta}]^{-1}\avg{\bm{\Theta}}\leq\frac{\Sigma}{2},\label{eq:mul.TUR}
\end{equation}
which relates the means and covariances of multiple currents.
For uncorrelated currents, i.e., $\mrm{Cov}[\Theta_i;\Theta_j]=\delta_{ij}\mrm{Var}[\Theta_i]$, one can obtain the following inequality from Eq.~\eqref{eq:mul.TUR}:
\begin{equation}
\sum_{i=1}^M\frac{\avg{\Theta_i}^2}{\mrm{Var}[\Theta_i]}\leq\frac{\Sigma}{2}.\label{eq:uncor.TUR}
\end{equation}
Equation~\eqref{eq:uncor.TUR} can be considered a generalization of Eq.~\eqref{eq:gen.TUR}.
We note that inequalities analogous to Eqs.~\eqref{eq:mul.TUR} and~\eqref{eq:uncor.TUR} have been derived for overdamped Langevin dynamics in Ref.~\cite{Dechant.2019.JPA}.
However, they do not hold for general underdamped systems, although our derived bounds are always satisfied.
\begin{figure}[t]
\centering
\includegraphics[width=8.5cm]{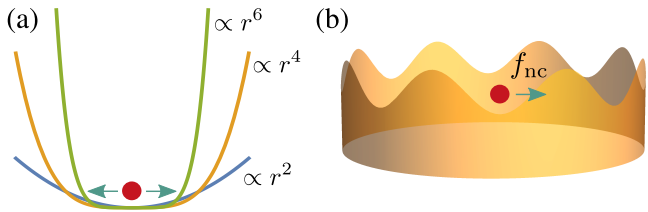}
\protect\caption{Schematic diagrams of systems used in simulations. (a) The particle is confined in a single-well potential $U(r)\propto r^{2n}$. There is no external force, and the system is in equilibrium. (b) The Brownian particle is driven out of equilibrium by a constant external force $f_{\rm nc}$ under a periodic potential $U(r)\propto\cos(2\pi nr/L)$.}\label{fig:models}
\end{figure}
\begin{figure}[t]
\centering
\includegraphics[width=8.5cm]{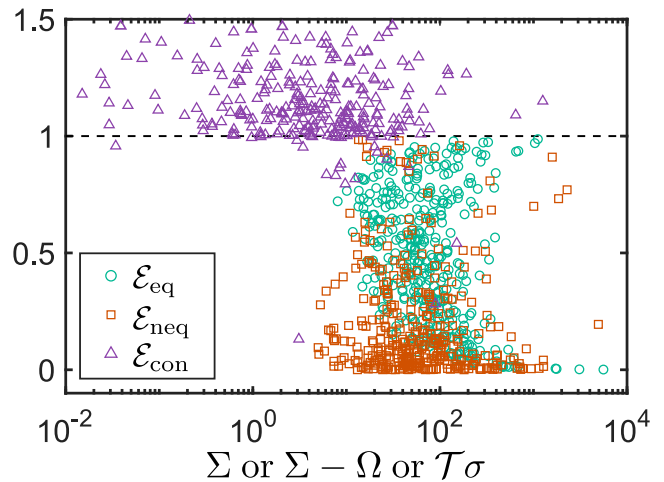}
\protect\caption{Numerical verification of the bounds. 
We randomly sample parameters and numerically solve the Langevin equation $10^{6}$ times with the time step $\Delta t=10^{-4}$.
The parameter ranges are $m,\gamma,\alpha,T\in[0.1,10]$ and $\mca{T}\in[1,10]$. 
The dashed line depicts each saturated case of the bounds.
For the single-well potential system, we calculate $\mca{E}_{\rm eq}\equiv8\avg{\Theta}^2/\bra{\Sigma\,\mrm{Var}[\Theta]}$, which should satisfy $\mca{E}_{\rm eq}\leq 1$, and plot $\mca{E}_{\rm eq}$ as a function of $\Sigma$ with green circles. 
All circular points are located below the line, thus empirically verifying the derived bound [Eq.~\eqref{eq:tig.TUR}].
Here, $n\in[1,3]$.
For the Brownian particle model, we evaluate $\mca{E}_{\rm neq}\equiv2\avg{\Theta}^2/\brab{\bra{\Sigma-\Omega}\mrm{Var}[\Theta]}$ since accurate calculation of $\Omega$ is difficult; $\mca{E}_{\rm neq}$ is plotted as a function of $\Sigma-\Omega$ with orange squares.
Since $\Omega$ is nonnegative, the derived bound [Eq.~\eqref{eq:gen.TUR}] is empirically verified if $\mca{E}_{\rm neq}\leq 1$ holds.
It is possible that $\mca{E}_{\rm neq}>1$ despite the derived bound being satisfied.
We also simultaneously calculate $\mca{E}_{\rm con}\equiv2\avg{\Theta}^2/\bra{\mca{T}\sigma\cdot\mrm{Var}[\Theta]}$, which corresponds to the original bound [Eq.~\eqref{eq:conv.TUR}], and plot it as a function of $\mca{T}\sigma$ with violet triangles.
$\mca{E}_{\rm con}>1$ indicates that the original bound is violated.
As seen, all square points lie below the dashed line and the derived bound is empirically verified.
By comparison, many triangular points are located above the line, thus implying that the original bound is violated.
Here, $L=1$, $f_{\rm nc}\in[0.1,10]$ and $n\in[1,10]$.}\label{fig:exa}
\end{figure}

\section{Applications}

\subsection{Illustrative examples}

We illustrate our results [Eqs.~\eqref{eq:gen.TUR} and \eqref{eq:tig.TUR}] with the aid of two systems.
We first consider an equilibrium system under the symmetric single-well potential $U(r)=\alpha r^{2n}/(2n)$ with $n$ as a positive integer, as illustrated in Fig.~\ref{fig:models}(a).
The total force acting upon the particle is $F(r)=-\alpha r^{2n-1}$.
For this system, the dynamical activity $\Upsilon$ and the boundary term $\Omega$ can be calculated analytically as
\begin{equation}
\Upsilon=\frac{\alpha^2}{T\gamma}\bra{\frac{\alpha}{2nT}}^{-2+1/n}\frac{G(2-\frac{1}{2n})}{G(\frac{1}{2n})}+\frac{\gamma}{m},~\Omega=4,
\end{equation}
where $G(z)=\int_0^{\infty}dx\,x^{z-1}e^{-x}$ is the gamma function.
Since the entropy production vanishes, $\Sigma$ can be expressed as $\Sigma=4\mca{T}\Upsilon+4$.
We consider current $\Theta[\bG]=\int_{0}^{\mca{T}}dt\,v^2$, which is proportional to the accumulated kinetic energy of the particle.
According to Eq.~\eqref{eq:tig.TUR}, the bound on this current fluctuation reads $\mrm{Var}[\Theta]/\avg{\Theta}^2\geq 8/\Sigma$, which is illustrated in Fig.~\ref{fig:exa}.
We find that the bound is satisfied for all selected parameter settings.

Next, we study a Brownian particle circulating on a ring of circumference $L$ under a periodic potential $U(r)=\alpha L/(2\pi n)\cos\bra{2\pi nr/L}$, with integer $n>0$ (see Fig.~\ref{fig:models}(b) for illustration).
The total force acting upon the particle is $F(r)=-\pp_{r}U(r)+f_{\rm nc}$, where $f_{\rm nc}$ is a constant external force that drives the particle out of equilibrium.
In the steady state, the entropy production rate becomes $\sigma=f_{\rm nc}\avg{v}/T$.
We validate the derived bound for current $\Theta[\bG]=\int_{0}^{\mca{T}}dt\,\dot{r}$, which corresponds to the accumulated distance traveled by the particle.
According to Eq.~\eqref{eq:gen.TUR}, we have $\mrm{Var}[\Theta]/\avg{\Theta}^2\geq2/\Sigma$.
This and the original bound are illustrated in Fig.~\ref{fig:exa}.
We find that the original bound can be violated; thus, Eq.~\eqref{eq:conv.TUR} does not hold for general underdamped systems.
In comparison, our bound is always satisfied and, thus, Eq.~\eqref{eq:gen.TUR} is empirically verified.

\subsection{TUR for active matter systems}

We consider a model system of active matter, namely, active Ornstein--Uhlenbeck particles (AOUPs), as has been studied in the literature \cite{Fodor.2016.PRL,Mandal.2017.PRL}.
The system contains $N$ self-propelled particles, which extract energy from the surrounding environment and exhibit self-induced motion.
The dynamics of the particle $i$ are governed by 
\begin{equation}
\dot{r}_i=-\mu\pp_{r_i}\Phi(\br)+\eta_i,~\tau\dot{\eta}_i=-\eta_i+\xi_i,\label{eq:act.mod}
\end{equation}
where $\mu$ is the mobility of particles, $\Phi(\br)$ is an interaction potential, $\xi_i$'s are zero-mean Gaussian white noises with properties $\avg{\xi_{i}(t)\xi_{j}(t')}=2D_i\delta_{ij}\delta(t-t')$, and $\eta_i$'s are Ornstein--Uhlenbeck processes with variances $\avg{\eta_i(0)\eta_j(t)}=D_i\delta_{ij}e^{-|t|/\tau}/\tau$.
Here, $\tau$ is the persistent time.
In the $\tau\to 0$ limit, $\eta_i$'s become white noises with delta-function variances and $r_i$'s become Markovian processes.
It is interesting that the force that drives the system out of equilibrium arises from the nonequilibrium environment.
To develop stochastic thermodynamics for the system, a mathematical mapping to an underdamped dynamics has been conducted by introducing velocity variables \cite{Fodor.2016.PRL,Nardini.2017.PRX,Mandal.2017.PRL}.
Specifically, taking the time derivative of Eq.~\eqref{eq:act.mod}, the system dynamics can be mapped to the following underdamped dynamics:
\begin{equation}
\dot{r}_i=v_i,~\tau \dot{v}_i=-v_i-\mu\bigg(1+\tau \sum_{k=1}^Nv_k\pp_{r_k}\bigg)\pp_{r_i}\Phi(\br)+\xi_i.\label{eq:map.und.dyn}
\end{equation}
By applying our approach to these underdamped dynamics, an uncertainty relation for an arbitrary observable $\Theta[\bG]=\int_{0}^{\mca{T}}dt\,\bL(\br)^\top\circ\dot{\br}$ can be obtained as follows:
\begin{equation}
\frac{\mrm{Var}[\Theta]}{\avg{\Theta}^2}\ge\frac{2}{\Sigma},
\end{equation}
where $\Sigma=\mca{T}(9\sigma+4\Upsilon_a)+\Omega$ and
\begin{equation}
\begin{aligned}
\Upsilon_a=&\sum_{i=1}^N\frac{1}{D_i}\bigg(\tau\mu\sum_{j=1}^N\avgl{v_iv_j\pp_{r_ir_j}^2\Phi(\br)}\\
&-2\avgl{\bram{\sum_{j=1}^Nv_j\bras{\delta_{ij}+\tau\mu\pp_{r_ir_j}^2\Phi(\br)}}^2}\\
&+\frac{3D_i}{\tau}\avg{1+\tau\mu\pp_{r_i}^2\Phi(\br)}\bigg).
\end{aligned}
\end{equation}
The detailed derivation of the bound is given in Appendix \ref{app:TUR.act.mat}.
We note that the definition of the total entropy production in the AOUPs system is not unique and depends on the chosen coarse-grained model.
The definition employed here is the same as in Ref.~\cite{Fodor.2016.PRL}.

\section{Conclusion}
In summary, we have derived the bounds on the current fluctuation in underdamped Langevin dynamics using information inequalities.
Our results indicated that the current fluctuation is constrained not only by the entropy production but also by the average dynamical activity.
The derived bound can be used as a tool for estimating thermodynamic quantities from the experimental data.
Deriving a bound for general currents of the form $\int\,dt\bras{\bL_1(\br,\bv)^\top\circ\dot{\br}+\bL_2(\br,\bv)^\top\circ\dot{\bv}}$ requires further investigation.

\section*{Acknowledgment}
This work was supported by Ministry of Education, Culture, Sports, Science and Technology (MEXT) KAKENHI Grants No. JP16K00325 and No. JP19K12153.

\appendix

\begin{widetext}
\section{Detailed calculations and derivations}

For the sake of convenience, we consider a more general system whose dynamics are described by
\begin{equation}
\dot{r}_i=v_i,~m_i\dot{v}_i=H_i(\br,\bv)+\xi_i,\label{eq:und.Lan}
\end{equation}
where $H_i(\br,\bv)=\sum_{j=1}^Nv_jG_{ij}(\br)+F_i(\br)$ and $\avg{\xi_i(t)\xi_j(t')}=2D_i\delta_{ij}\delta(t-t')$.
The system employed in the paper can be obtained by substituting
\begin{equation}
G_{ij}(\br)\leftarrow-\delta_{ij}\gamma_i,~ D_i\leftarrow T_i\gamma_i.
\end{equation}

\subsection{Calculation of entropy production rate}\label{app:ent.cal}

Following Ref.~\cite{Spinney.2012.PRE}, we calculate the entropy production rate.
Let us consider an infinitesimal time interval $[t,t+dt]$, in which $\br\equiv\br(t),\bv\equiv\bv(t),\br'\equiv\br(t+dt)$, and $\bv'\equiv\bv(t+dt)$.
From the definition of the entropy production in Eq.~\eqref{eq:ent.decomp}, the entropy production in this time interval is given by
\begin{equation}
d\Delta s_{\rm tot}=-d(\ln P(\xs,t))+\ln\frac{\bms{P}[\br',\bv',t+dt|\br,\bv,t]}{\bms{P}[\br,-\bv,t+dt|\br',-\bv',t]}.\label{eq:inc.tot.ent}
\end{equation}
By using the short time propagator, we can express the transition probability as
\begin{equation}
\bms{P}[\br',\bv',t+dt|\br,\bv,t]=\prod_{i=1}^N\frac{m_i}{\sqrt{4\pi D_idt}}\exp\bras{-\frac{\bra{m_idv_i-H_i(\br,\bv)dt}^2}{4D_idt}},\label{eq:tran.prob.forw}
\end{equation}
where $dv_i=v_i'-v_i$.
The force $H_i(\xs)$ can be decomposed into reversible and irreversible parts as $H_i(\xs)=H_i^{\rm ir}(\xs)+H_i^{\rm rev}(\xs)$, where $H_i^{\rm ir}(\xs)=\sum_{j=1}^Nv_jG_{ij}(\br)$ and $H_i^{\rm rev}(\xs)=F_i(\br)$.
Analogously, we also have
\begin{equation}
\bms{P}[\br,-\bv,t+dt|\br',-\bv',t]=\prod_{i=1}^N\frac{m_i}{\sqrt{4\pi D_idt}}\exp\bras{-\frac{\bra{m_idv_i-\bras{-H_i^{\rm ir}(\br,\bv)+H_i^{\rm rev}(\br,\bv)}dt}^2}{4D_idt}-\frac{1}{m_i}\pp_{v_i}H_i^{\rm ir}(\xs)dt}.\label{eq:tran.prob.back}
\end{equation}
We note that in the case of additive noise, the discretization schemes in the forward and backward paths are independent.
However, for multiplicative noise, there is a constraint on the discretization.
That is, if the evaluation points in the forward path are $a\br'+(1-a)\br$ and $a\bv'+(1-a)\bv$, then in the backward path, they should be $b\br+(1-b)\br'$ and $b\bv+(1-b)\bv'$, where $b=1-a$.
Here, we have employed the discretization with $a=0$ and $b=1$.
Using Eqs.~\eqref{eq:tran.prob.forw} and \eqref{eq:tran.prob.back}, we obtain
\begin{equation}
\ln\frac{\bms{P}[\br',\bv',t+dt|\br,\bv,t]}{\bms{P}[\br,-\bv,t+dt|\br',-\bv',t]}=\sum_{i=1}^N\bras{\frac{m_i}{D_i}H_i^{\rm ir}(\xs)dv_i-\frac{1}{D_i}H_i^{\rm ir}(\xs)H_i^{\rm rev}(\xs)dt+\frac{1}{m_i}\pp_{v_i}H_i^{\rm ir}(\xs)dt}.\label{eq:inc.med.ent}
\end{equation}
The change in the system entropy is written in Ito rules as
\begin{equation}
\begin{aligned}
-d(\ln P(\xs,t))=&-\frac{1}{P(\xs,t)}\pp_{t}P(\xs,t)dt-\frac{1}{P(\xs,t)}\sum_{i=1}^N\pp_{r_i}P(\xs,t)dr_i-\frac{1}{P(\xs,t)}\sum_{i=1}^N\pp_{v_i}P(\xs,t)dv_i\\
&-\sum_{i=1}^N\frac{D_i}{m_i^2P(\xs,t)}\bras{\pp_{v_i}^2P(\xs,t)-\frac{1}{P(\xs,t)}\bra{\pp_{v_i}P(\xs,t)}^2}.\label{eq:inc.sys.ent}
\end{aligned}	
\end{equation}
Plugging Eqs.~\eqref{eq:inc.med.ent} and \eqref{eq:inc.sys.ent} into Eq.~\eqref{eq:inc.tot.ent}, we obtain
\begin{equation}\label{eq:tot.ent.evo}
\begin{aligned}
d\Delta s_{\rm tot}=&\sum_{i=1}^N\bras{\frac{m_i}{D_i}H_i^{\rm ir}(\xs)dv_i-\frac{1}{D_i}H_i^{\rm ir}(\xs)H_i^{\rm rev}(\xs)dt+\frac{1}{m_i}\pp_{v_i}H_i^{\rm ir}(\xs)dt}\\
&-\frac{1}{P(\xs,t)}\pp_{t}P(\xs,t)dt-\frac{1}{P(\xs,t)}\sum_{i=1}^N\pp_{r_i}P(\xs,t)dr_i-\frac{1}{P(\xs,t)}\sum_{i=1}^N\pp_{v_i}P(\xs,t)dv_i\\
&-\sum_{i=1}^N\frac{D_i}{m_i^2P(\xs,t)}\bras{\pp_{v_i}^2P(\xs,t)-\frac{1}{P(\xs,t)}\bra{\pp_{v_i}P(\xs,t)}^2}.
\end{aligned}
\end{equation}
In the steady state, i.e., $P(\xs,t)=\pss(\xs)$, the average entropy production can be calculated as
\begin{equation}
\avg{d\Delta s_{\rm tot}}=\iint\dxs\pss(\xs)\avg{d\Delta s_{\rm tot}|\xs},
\end{equation}
where $\avg{d\Delta s_{\rm tot}|\xs}$ can be evaluated by replacing terms in $d\Delta s_{\rm tot}$ such that $dr_i=v_idt$ and $m_idv_i=H_i(\xs)dt$.
Then, we obtain
\begin{equation}
\begin{aligned}
\avg{d\Delta s_{\rm tot}}=&\sum_{i=1}^N\iint\dxs\bigg[\frac{1}{D_i}\pss(\xs)H_i^{\rm ir}(\xs)^2+\frac{\pss(\xs)}{m_i}\pp_{v_i}H_i^{\rm ir}(\xs)-v_i\pp_{r_i}\pss(\xs)-\frac{H_i(\xs)}{m_i}\pp_{v_i}\pss(\xs)\\
&-\frac{D_i}{m_i^2}\pp_{v_i}^2\pss(\xs)+\frac{D_i}{m_i^2\pss(\xs)}\bra{\pp_{v_i}\pss(\xs)}^2\bigg]dt\\
=&\sum_{i=1}^N{\iint\dxs\frac{\bra{H_i^{\rm ir}(\xs)\pss(\xs)-D_i/m_i\pp_{v_i}\pss(\xs)}^2}{D_i\pss(\xs)}}dt\\
=&\sum_{i=1}^N{\iint\dxs\frac{m_i^2}{D_i}\frac{J_{v_i}^{\rm ir}(\xs)^2}{\pss(\xs)}}dt,
\end{aligned}
\end{equation}
where $J_{v_i}^{\rm ir}(\xs)=1/m_i\bras{H_i^{\rm ir}(\xs)-D_i/m_i\pp_{v_i}}\pss(\xs)$.
Finally, the entropy production rate is given by
\begin{equation}
\sigma\equiv{\frac{\avg{d\Delta s_{\rm tot}}}{dt}}=\sum_{i=1}^N\iint\dxs\frac{m_i^2}{D_i}\frac{J_{v_i}^{\rm ir}(\xs)^2}{\pss(\xs)}.
\end{equation}
Now, by setting
\begin{equation}
G_{ij}(\br)\leftarrow -\delta_{ij}\gamma_i,\quad D_i\leftarrow T_i\gamma_i,
\end{equation}
the irreversible current becomes $J_{v_i}^{\rm ir}(\xs)=-1/m_i\bras{\gamma_iv_i+T_i\gamma_i/m_i\pp_{v_i}}\pss(\xs)$, and the entropy production rate reads
\begin{equation}
\sigma=\sum_{i=1}^N\iint\dxs\frac{m_i^2}{T_i\gamma_i}\frac{J_{v_i}^{\rm ir}(\xs)^2}{\pss(\xs)}.
\end{equation}

\subsection{Path integral}\label{app:pat.int}

We discretize time by dividing the interval $[0,\mca{T}]$ into $K$ equipartitioned intervals with a time step $\Delta t$, where $\mca{T}=K\Delta t$, $t^{k}\equiv k\Delta t$, $r^{k}_i\equiv r_i(t^{k})$, and $v_i^{k}\equiv v_i(t^{k})$ (superscripts denote points in a temporal sequence). Discretization of Eq.~\eqref{eq:und.Lan} yields
\begin{equation}
\begin{aligned}
r_i^{k+1}-r_i^{k}&=v_i^k\Delta t,\\
m_i(v_i^{k+1}-v_i^{k})&=H_i(r_i^k,v_i^k)\Delta t+\Delta w_i^k,\label{eq:discretized_Langevin}
\end{aligned}
\end{equation}
where $\Delta w_i^{k} \equiv w_i^{k+1}-w_i^{k}=w_i(t^{k+1})-w_i(t^k)$ is a Wiener process with the following properties:
\begin{equation}
\avg{\Delta w_i^{k}}=0,\quad\avg{\Delta w_i^{k}\Delta w_{i^{\prime}}^{k^{\prime}}}=2\delta_{ii^{\prime}}\delta_{kk^{\prime}}D_i\Delta t.
\end{equation}
A stochastic trajectory $\bm{\Gamma}\equiv\left[\br^0,\bv^0,\br^1,\bv^{1},\br^2,\bv^{2},\dots,\br^K,\bv^{K}\right]$ is specified given $\bm{\mca{W}}\equiv\left[\Delta\bm{w}^{0},\Delta\bm{w}^{1},\dots,\Delta\bm{w}^{K-1}\right]$ and $(\br^0,\bv^0)$.
The probability density function of the Wiener processes $\Delta\bm{w}^{k}$ is given by Eq.~\eqref{eq:w_PDF_def}:
\begin{equation}
\bms{P}[\bm{\mca{W}}]=\prod_{i=1}^N\prod_{k=0}^{K-1}P(\Delta w_i^{k})=\prod_{i=1}^N\prod_{k=0}^{K-1}\frac{1}{\sqrt{4\pi D_i\Delta t}}\exp\left[-\frac{(\Delta w_i^{k})^{2}}{4D_i\Delta t}\right].\label{eq:w_PDF_def}
\end{equation}
Let us change the variables in Eq.~\eqref{eq:w_PDF_def} from $\bm{\mca{W}}=\left[\Delta \bm{w}^{0},\Delta \bm{w}^{1},\dots,\Delta \bm{w}^{K-1}\right]$ to $\bm{\mca{V}}=\left[\bv^{1},\bv^{2},\dots,\bv^{K}\right]$.
From Eq.~\eqref{eq:discretized_Langevin}, the determinant of the Jacobian matrix is
\begin{equation}
\left|\frac{\pp(\bv^{1},\dots,\bv^{K})}{\pp(\Delta \bm{w}^{0},\dots,\Delta \bm{w}^{K-1})}\right|=\prod_{i=1}^N\prod_{k=0}^{K-1}\frac{1}{m_i},\label{eq:Jacobian_def}
\end{equation}
given that the determinant of triangular matrices is a product of their diagonal elements. 
Using Eqs.~\eqref{eq:discretized_Langevin}, \eqref{eq:w_PDF_def}, and \eqref{eq:Jacobian_def}, we obtain
\begin{equation}
\bms{P}[\bm{\Gamma}|\br^{0},\bv^0]=\bra{\prod_{i=1}^N\prod_{k=0}^{K-1}\frac{m_i}{\sqrt{4\pi D_i\Delta t}}}\exp\bras{-\sum_{i=1}^N\frac{\Delta t}{4D_i}\sum_{k=0}^{K-1}\bra{\frac{m_i(v_i^{k+1}-v_i^{k})}{\Delta t}-H_i(r_i^{k},v_i^k)}^{2}}.\label{eq:PX_pathintegral_def}
\end{equation}
In the limit $K\to\infty$, we obtain the path integral in Eq.~\eqref{eq:path.int.pre}.
\begin{equation}
\bms{P}[\bm{\Gamma}|\br^{0},\bv^0]=\mca{N}\prod_{i=1}^N\exp{\bras{-\frac{1}{4D_i}\int_0^{\mca{T}}dt\bra{m_i\dot{v}_i-H_i(\xs)}^2}}\label{eq:path.int.pre}
\end{equation}
By substituting such that $D_i=T_i\gamma_i$ and $H_i(\xs)=-\gamma_iv_i+F_i(\br)$, we have
\begin{equation}
\bms{P}[\bm{\Gamma}|\br^{0},\bv^0]=\mca{N}\prod_{i=1}^N\exp{\bras{-\frac{1}{4T_i\gamma_i}\int_0^{\mca{T}}dt\bra{m_i\dot{v}_i+\gamma_iv_i-F_i(\br)}^2}}.
\end{equation}
The path integral in Eq.~\eqref{eq:path.int.pre} is a continuous limit of the Ito sum, which can be transformed to a Stratonovich integral via the following rule:
\begin{equation}
\int_0^{\mca{T}} dt~\dot{v}_i\cdot H_i(\xs)=\int_0^{\mca{T}} dt~\dot{v}_i\circ H_i(\xs)-\frac{D_i}{m_i^2}\int_0^{\mca{T}} dt~\pp_{v_i}H_i(\xs),
\end{equation}
where $\cdot$ and $\circ$ denote the Ito and Stratonovich products, respectively.
The path integral using the mid-point discretization is then expressed as
\begin{equation}
\bms{P}[\bm{\Gamma}|\br^{0},\bv^0]=\mca{N}\prod_{i=1}^N\exp{\bras{-\frac{1}{4D_i}\int_0^{\mca{T}}dt\brab{\bra{m_i\dot{v}_i-H_i(\xs)}^2+\frac{2D_i}{m_i}\pp_{v_i}H_i(\xs)}}}.\label{eq:path.int.mid}
\end{equation}
For one-dimensional underdamped dynamics
\begin{equation}
\dot{r}=v,\quad m\dot{v}=-\gamma v+F(r)+\xi(t),
\end{equation}
the path probability written as the Stratonovich integral becomes
\begin{equation}
\bms{P}[\bm{\Gamma}|r^{0},v^0]=\mca{N}\exp\bra{\frac{\mca{T}\gamma}{2m}}\exp{\bras{-\frac{1}{4T\gamma}\int_0^{\mca{T}}dt\bra{m\dot{v}+\gamma v-F(r)}^2}},
\end{equation}
which is consistent with the result in Ref.~\cite{Rosinberg.2016.EPL}.

\subsection{Bound on the fluctuation of currents}\label{app:TUR.deriv}

Let $\pss(\xs)$ be the steady-state distribution of the original dynamics.
Now, we consider the following auxiliary dynamics
\begin{equation}
\dot{r}_i=v_i,~m_i\dot{v}_i=H_{i,\theta}(\xs)+\xi_i,
\end{equation}
where
\begin{equation}
H_{i,\theta}(\xs)=(1+\theta)\sum_{j=1}^Nv_jG_{ij}(\br)+(1+\theta)^2F_i(\br)+\frac{D_i}{m_i}(1-(1+\theta)^3)\frac{\pp_{v_i}\pss(\mxs)}{\pss(\mxs)}.
\end{equation}
We note that when $\theta=0$, $H_{i,\theta}(\xs)=\sum_{j=1}^Nv_jG_{ij}(\br)+F_i(\br)$ and the auxiliary dynamics become the original ones.
The corresponding Kramers equation of this dynamics is
\begin{equation}
\pp_tP_\theta(\xs,t)=\sum_{i=1}^{N}\bras{-\pp_{r_i}J_{r_i,\theta}(\xs,t)-\pp_{v_i}J_{v_i,\theta}(\xs,t)},
\end{equation}
where
\begin{equation}
J_{r_i,\theta}(\xs,t)=v_iP_\theta(\xs,t),~ J_{v_i,\theta}(\xs,t)=1/m_i\bras{H_{i,\theta}(\xs)-D_i/m_i\pp_{v_i}}P_\theta(\xs,t).
\end{equation}
It can be easily proven that the steady-state distribution of the auxiliary dynamics is $\mpss(\xs)=\pss(\mxs)/(1+\theta)^{N}$.
Since $\avg{\Theta}_\theta=(1+\theta)\avg{\Theta}\Rightarrow\pp_\theta\avg{\Theta}_\theta=\avg{\Theta}$, the Cram{\'e}r-Rao inequality when letting $\theta=0$ reads
\begin{equation}
\frac{\mrm{Var}[\Theta]}{\avg{\Theta}^2}\geq\frac{1}{\mca{I}(0)}.\label{eq:raw.TUR}
\end{equation}
The Fisher information can be calculated as
\begin{align}
\mca{I}(0)=&-\avgl{\pp_\theta^2\ln P_{\theta}^{\rm ss}(\br^0,\bv^0)}_{\theta=0}+\frac{1}{2}\avgl{\int_0^{\mca{T}}dt\,\sum_{i=1}^N\frac{1}{D_i}\bra{\pp_\theta H_{i,\theta}(\xs)}^2}_{\theta=0}\\
=&\avgl{\bras{\pp_\theta\ln P_\theta^{\rm ss}(\br^0,\bv^0)}^2}_{\theta=0}+\frac{\mca{T}}{2}\sum_{i=1}^N\frac{1}{D_i}\avgl{\bra{\pp_{\theta}H_{i,\theta}(\xs)}^2}_{\theta=0}\\
=&\avgl{\bras{\pp_\theta\ln P_\theta^{\rm ss}(\br^0,\bv^0)}^2}_{\theta=0}+\frac{\mca{T}}{2}\sum_{i=1}^N\frac{1}{D_i}\avgl{\bram{\sum_{j=1}^Nv_jG_{ij}(\br)+2F_i(\br)-3\frac{D_i}{m_i}\frac{\pp_{v_i}\pss(\xs)}{\pss(\xs)}}^2}\\
=&\avgl{\bras{\pp_\theta\ln P_\theta^{\rm ss}(\br^0,\bv^0)}^2}_{\theta=0}\nonumber\\
&+\frac{\mca{T}}{2}\sum_{i=1}^N\frac{1}{D_i}\avgl{4\bram{F_i(\br)-\sum_{j=1}^Nv_jG_{ij}(\br)}^2+9m_i^2\bra{\frac{J_{v_i}^{\rm ir}(\xs)}{\pss(\xs)}}^2+12m_i\frac{J_{v_i}^{\rm ir}(\xs)}{\pss(\xs)}\bram{F_i(\br)-\sum_{j=1}^Nv_jG_{ij}(\br)}}.\label{eq:last.trans.I}
\end{align}
In Eq.~\eqref{eq:last.trans.I}, we used the relation that $J_{v_i}^{\rm ir}(\xs)=1/m_i\bras{\sum_{j=1}^Nv_jG_{ij}(\br)\pss(\xs)-D_i/m_i\pp_{v_i}\pss(\xs)}$.
Now, we transform each term in $\mca{I}(0)$.
The first term, which is a boundary value, can be evaluated as
\begin{equation}
\avgl{\bras{\pp_\theta\ln P_\theta^{\rm ss}(\br^0,\bv^0)}^2}_{\theta=0}=\bigg\langle{\Big(\sum_{i=1}^{N}v_i\pp_{v_i}\pss(\xs)/\pss(\xs)\Big)^2}\bigg\rangle-N^2.\label{eq:first.term.I0}
\end{equation}
The second term in $\mca{I}(0)$ can be transformed as
\begin{equation}
\avgl{\bram{F_i(\br)-\sum_{j=1}^Nv_jG_{ij}(\br)}^2}=\avg{F_i(\br)^2}+\avgl{\bram{\sum_{j=1}^Nv_jG_{ij}(\br)}^2}-2\sum_{j=1}^N\avg{v_jF_i(\br)G_{ij}(\br)}.\label{eq:second.term.I0}
\end{equation}
The third term is equal to the entropy production rate
\begin{equation}
\sum_{i=1}^N\frac{m_i^2}{D_i}\avgl{\bra{\frac{J_{v_i}^{\rm ir}(\xs)}{\pss(\xs)}}^2}=\sum_{i=1}^N\iint\dxs\frac{m_i^2}{D_i}\frac{J^{\rm ir}_{v_i}(\xs)^2}{\pss(\xs)}=\sigma.\label{eq:third.term.I0}
\end{equation}
The last term can be calculated as
\begin{equation}
\avgl{\frac{m_iJ_{v_i}^{\rm ir}(\xs)}{\pss(\xs)}\bram{F_i(\br)-\sum_{j=1}^Nv_jG_{ij}(\br)}}=\sum_{j=1}^N\avg{v_jF_i(\br)G_{ij}(\br)}-\avgl{\bram{\sum_{j=1}^Nv_jG_{ij}(\br)}^2}-\frac{D_i}{m_i}\avg{G_{ii}(\br)}.\label{eq:last.term.I0}
\end{equation}
Collecting the terms in Eqs.~\eqref{eq:first.term.I0}--\eqref{eq:last.term.I0}, we can rewritte $\mca{I}(0)$ as
\begin{equation}
\mca{I}(0)=\frac{1}{2}\bras{\mca{T}\bra{9\sigma+4\Upsilon}+\Omega},
\end{equation}
where
\begin{align}
\Upsilon&=\sum_{i=1}^N\frac{1}{D_i}\brah{\avg{F_i(\br)^2}-2\avgl{\bram{\sum_{j=1}^Nv_jG_{ij}(\br)}^2}+\sum_{j=1}^N\avg{v_jF_i(\br)G_{ij}(\br)}-\frac{3D_i}{m_i}\avg{G_{ii}(\br)}},\label{eq:ups}\\
\Omega&=2\bigg\langle{\Big(\sum_{i=1}^{N}v_i\pp_{v_i}\pss(\xs)/\pss(\xs)\Big)^2}\bigg\rangle-2N^2.\label{eq:omgs}
\end{align}
The bound in Eq.~\eqref{eq:raw.TUR} then becomes
\begin{equation}
\frac{\mrm{Var}[\Theta]}{\avg{\Theta}^2}\geq\frac{2}{\mca{T}(9\sigma+4\Upsilon)+\Omega}.\label{eq:raw.TUR.2}
\end{equation}
Now, by substituting $G_{ij}(\br)=-\delta_{ij}\gamma_i$ and $D_i=T_i\gamma_i$ into $\Upsilon$, we have
\begin{equation}
\Upsilon=\sum_{i=1}^N\frac{1}{T_i\gamma_i}\brah{\avg{F_i(\br)^2}-2\gamma_i^2\avg{v_i^2}-\gamma_i\avg{v_iF_i(\br)}+\frac{3T_i\gamma_i^2}{m_i}}.
\end{equation}
Since
\begin{equation}
\avg{v_iF_i(\br)}=\gamma_i\avg{v_i^2}-\frac{T_i\gamma_i}{m_i},
\end{equation}
$\Upsilon$ can be further simplified as
\begin{equation}
\Upsilon=\sum_{i=1}^N\brah{\frac{1}{T_i\gamma_i}\avg{F_i(\br)^2}-3\frac{\gamma_i}{T_i}\avg{v_i^2}+4\frac{\gamma_i}{m_i}}.
\end{equation}
Finally, we obtain the bound
\begin{equation}
\frac{\mrm{Var}[\Theta]}{\avg{\Theta}^2}\ge\frac{2}{\Sigma},
\end{equation}
where $\Sigma=\mca{T}(9\sigma+4\Upsilon)+\Omega$.

\subsection{Equality condition of the derived bound}\label{app:equ.cond}

The equality condition of the derived bound is that the following relation,
\begin{equation}
\left.\pp_\theta\ln\bms{P}_\theta[\bm{\Gamma}]\right|_{\theta=0}=\mu\bras{\Theta[\bm{\Gamma}]-\psi(0)},\label{eq:TUR.equ.con}
\end{equation}
holds for an arbitrary trajectory $\bm{\Gamma}$.
Using the formula of the path integral, we have
\begin{equation}
\begin{aligned}
\left.\pp_\theta\ln\bms{P}_\theta[\bm{\Gamma}]\right|_{\theta=0}=&\left.\pp_\theta\ln P_\theta^{\rm ss}(\br^0,\bv^0)\right|_{\theta=0}+\left.\sum_{i=1}^{N}\frac{1}{2D_i}\int_{0}^{\mca{T}}dt\,\pp_{\theta}H_{i,\theta}\bra{\dot{v}_i-H_{i,\theta}}\right|_{\theta=0}\\
=&-N-\sum_{i=1}^{N}\frac{v^0_{i}\pp_{v_i}\pss(\br^0,\bv^0)}{\pss(\br^0,\bv^0)}\\
&+\sum_{i=1}^{N}\frac{1}{2D_i}\int_{0}^{\mca{T}}dt\,\bra{-\gamma_iv_i+2F_i(\br)-\frac{3D_i}{m_i}\frac{\pp_{v_i}\pss(\xs)}{\pss(\xs)}}\cdot\bra{m_i\dot{v}_i+\gamma_iv_i-F_i(\br)}\\
=&-N-\sum_{i=1}^{N}\frac{v^0_{i}\pp_{v_i}\pss(\br^0,\bv^0)}{\pss(\br^0,\bv^0)}+\sum_{i=1}^{N}\frac{1}{2m_i}\int_{0}^{\mca{T}}dt\,\bra{\gamma_i+\frac{3D_i}{m_i}\pp_{v_i}\bra{\frac{\pp_{v_i}\pss(\xs)}{\pss(\xs)}}}\\
&+\sum_{i=1}^{N}\frac{1}{2D_i}\int_{0}^{\mca{T}}dt\,\bra{-\gamma_iv_i+2F_i(\br)-\frac{3D_i}{m_i}\frac{\pp_{v_i}\pss(\xs)}{\pss(\xs)}}\circ\bra{m_i\dot{v}_i+\gamma_iv_i-F_i(\br)}.
\end{aligned}
\end{equation}
Therefore, Eq.~\eqref{eq:TUR.equ.con} is equivalent with
\begin{equation}\label{eq:TUR.equ.con2}
\begin{aligned}
&-N-\sum_{i=1}^{N}\frac{v^0_{i}\pp_{v_i}\pss(\br^0,\bv^0)}{\pss(\br^0,\bv^0)}+\sum_{i=1}^{N}\frac{1}{2D_i}\int_{0}^{\mca{T}}dt\,\bra{-\gamma_iv_i+2F_i(\br)-\frac{3D_i}{m_i}\frac{\pp_{v_i}\pss(\xs)}{\pss(\xs)}}\circ\bra{m_i\dot{v}_i+\gamma_iv_i-F_i(\br)}\\
&=\mu\bras{\int_0^{\mca{T}}dt\,\bm{\Lambda}(\br)^\top\circ\dot{\br}-\psi(0)}-\sum_{i=1}^{N}\frac{1}{2m_i}\int_{0}^{\mca{T}}dt\,\bra{\gamma_i+\frac{3D_i}{m_i}\pp_{v_i}\bra{\frac{\pp_{v_i}\pss(\xs)}{\pss(\xs)}}}.
\end{aligned}
\end{equation}
Since the left-hand side of Eq.~\eqref{eq:TUR.equ.con2} contains $\dot{\bv}$ while the right-hand side does not, Eq.~\eqref{eq:TUR.equ.con2} only holds for all trajectories when the term $\dot{\bv}$ disappears, i.e., when 
\begin{equation}
-\gamma_iv_i+2F_i(\br)-3\frac{D_i}{m_i}\frac{\pp_{v_i}\pss(\xs)}{\pss(\xs)}=0\label{eq:new.TUR.equ.con.1}
\end{equation}
for all $\xs$.
Consequently, Eq.~\eqref{eq:TUR.equ.con2} becomes
\begin{equation}
-N-\sum_{i=1}^{N}\frac{v^0_{i}\pp_{v_i}\pss(\br^0,\bv^0)}{\pss(\br^0,\bv^0)}=\mu\bras{\int_0^{\mca{T}}dt\,\bm{\Lambda}(\br)^\top\circ\dot{\br}-\psi(0)}-\sum_{i=1}^{N}\frac{1}{2m_i}\int_{0}^{\mca{T}}dt\,\bra{\gamma_i+\frac{3D_i}{m_i}\pp_{v_i}\bra{\frac{\pp_{v_i}\pss(\xs)}{\pss(\xs)}}}.\label{eq:TUR.equ.con3}
\end{equation}
Now, we take the partial derivative of both sides of Eq.~\eqref{eq:new.TUR.equ.con.1} with respect to $v_i$ to obtain
\begin{equation}
\gamma_i+\frac{3D_i}{m_i}\pp_{v_i}\bra{\frac{\pp_{v_i}\pss(\xs)}{\pss(\xs)}}=0.\label{eq:TUR.equ.con4}
\end{equation}
Using Eq.~\eqref{eq:TUR.equ.con4}, we can simplify Eq.~\eqref{eq:TUR.equ.con3} to be
\begin{equation}
-N-\sum_{i=1}^{N}\frac{m_i}{3D_i}v^0_{i}\bra{-\gamma_iv^0_{i}+2F_i(\br^0)}=\mu\bras{\int_0^{\mca{T}}dt\,\bm{\Lambda}(\br)^\top\circ\dot{\br}-\psi(0)}\label{eq:TUR.equ.con5}.
\end{equation}
As can be seen, the term in the left-hand side of Eq.~\eqref{eq:TUR.equ.con5} is only dependent on the initial point of the trajectory, $(\br^0,\bv^0)$, while the term in the right-hand side depends entirely on the trajectory $\bm{\Gamma}$.
Therefore, Eq.~\eqref{eq:TUR.equ.con5} does not hold for all trajectories, thus implying that the equality condition of the derived bound cannot be attained.

\subsection{Multidimensional TUR}\label{app:mul.TUR}

\subsubsection{Derivation of Eq.~\eqref{eq:mul.CRI}}\label{app:eq.mul.CRI}
Following Ref.~\cite{Bos.2007}, here we show a proof of Eq.~\eqref{eq:mul.CRI}.
First, noticing that
\begin{equation}
\label{eq:mul.TUR.1}
\pp_\theta\avg{\bm{\Theta}}_\theta=\pp_\theta\int\mca{D}\bG\,\bms{P}_\theta[\bG]\bT[\bG]=\int\mca{D}\bG\,\bms{P}_\theta[\bG]\bT[\bG]\pp_\theta\ln\bms{P}_\theta[\bG]=\mbb{E}[\bT[\bG]\pp_\theta\ln\bms{P}_\theta[\bG]],
\end{equation}
where $\mbb{E}[\cdot]$ denotes the average taken over all possible trajectories.
Since $\mbb{E}[\pp_\theta\ln\bms{P}_\theta[\bG]]=\pp_\theta\int\mca{D}\bG\,\bms{P}_\theta[\bG]=0$, Eq.~\eqref{eq:mul.TUR.1} can be written
\begin{equation}
\pp_\theta\avg{\bm{\Theta}}_\theta=\mrm{Cov}[\bT;\,\pp_\theta\ln\bms{P}_\theta].
\end{equation}
Now, consider the following vector:
\begin{equation}
\begin{pmatrix}
	\bT[\bG]\\
	\pp_\theta\ln\bms{P}_\theta[\bG]
\end{pmatrix}\in\mbb{R}^{(M+1)\times 1},
\end{equation}
whose covariance matrix,
\begin{equation}
\begin{pmatrix}
	\mrm{Cov}[\bT] & \mrm{Cov}[\bT;\,\pp_\theta\ln\bms{P}_\theta]\\
	\mrm{Cov}[\bT;\,\pp_\theta\ln\bms{P}_\theta]^\top & \mrm{Cov}[\pp_\theta\ln\bms{P}_\theta]
\end{pmatrix},
\end{equation}
is positive semi-definte and is equal to
\begin{equation}
\begin{pmatrix}
	\mrm{Cov}[\bT] & \pp_\theta\avg{\bm{\Theta}}_\theta\\
	\pp_\theta\avg{\bm{\Theta}}_\theta^\top & \mca{I}(\theta)
\end{pmatrix}.
\end{equation}
Using the fact that if a matrix $\bm{X}$ is positive semi-definte, then so is $\bm{Y}^\top\bm{X}\bm{Y}$, we have that
\begin{equation}\label{eq:tmp.mat}
\begin{pmatrix}
\bm{I} & -\mca{I}(\theta)^{-1}\pp_\theta\avg{\bT}_\theta
\end{pmatrix}
\begin{pmatrix}
	\mrm{Cov}[\bT] & \pp_\theta\avg{\bm{\Theta}}_\theta\\
	\pp_\theta\avg{\bm{\Theta}}_\theta^\top & \mca{I}(\theta)
\end{pmatrix}
\begin{pmatrix}
\bm{I}\\
-\mca{I}(\theta)^{-1}\pp_\theta\avg{\bT}_\theta^\top
\end{pmatrix}
\end{equation}
is also positive semi-definite, where $\bm{I}\in\mbb{R}^{M\times M}$ is the identity matrix.
After performing some matrix multiplications in Eq.~\eqref{eq:tmp.mat}, we obtain that the matrix
\begin{equation}
\mrm{Cov}[\bT]-\mca{I}(\theta)^{-1}\pp_\theta\avg{\bT}_\theta\pp_\theta\avg{\bT}_\theta^\top
\end{equation}
is positive semi-definite, i.e., $\mrm{Cov}[\bT]\succeq\mca{I}(\theta)^{-1}\pp_\theta\avg{\bT}_\theta\pp_\theta\avg{\bT}_\theta^\top$.

\subsubsection{Derivation of Eq.~\eqref{eq:mul.TUR}}\label{app:eq.mul.TUR}

By substituting $\bm{x}=\avg{\bm{\Theta}}$ into $\bm{x}^\top\bra{\mrm{Cov}[\bm{\Theta}]-2\avg{\bm{\Theta}}\avg{\bm{\Theta}}^\top/\Sigma}^{-1}\bm{x}\geq 0$, we have
\begin{equation}
\avg{\bm{\Theta}}^\top\bra{\mrm{Cov}[\bm{\Theta}]-2\avg{\bm{\Theta}}\avg{\bm{\Theta}}^\top/\Sigma}^{-1}\avg{\bm{\Theta}}\geq 0.\label{eq:mul.TUR.der}
\end{equation}
Using the Sherman--Morrison formula,
\begin{equation}
(\bm{A}+\bm{u}\bm{v}^\top)^{-1}=\bm{A}^{-1}-\frac{\bm{A}^{-1}\bm{u}\bm{v}^\top \bm{A}^{-1}}{1+\bm{v}^\top \bm{A}^{-1}\bm{u}},
\end{equation}
where $\bm{A}\in\mbb{R}^{n\times n}$ and $\bm{u},\bm{v}\in\mbb{R}^{n}$, Eq.~\eqref{eq:mul.TUR.der} can be transformed as follows:
\begin{equation}
\begin{aligned}
\avg{\bm{\Theta}}^\top\bra{\mrm{Cov}[\bm{\Theta}]-2\avg{\bm{\Theta}}\avg{\bm{\Theta}}^\top/\Sigma}^{-1}\avg{\bm{\Theta}}=&\avg{\bm{\Theta}}^\top\bra{\mrm{Cov}[\bm{\Theta}]^{-1}+\frac{2\mrm{Cov}[\bm{\Theta}]^{-1}\avg{\bm{\Theta}}\avg{\bm{\Theta}}^\top\mrm{Cov}[\bm{\Theta}]^{-1}/\Sigma}{1-2\avg{\bm{\Theta}}^\top\mrm{Cov}[\bm{\Theta}]^{-1}\avg{\bm{\Theta}}/\Sigma}}\avg{\bm{\Theta}}\\
=&z+\frac{2z^2/\Sigma}{1-2z/\Sigma}\ge 0,\quad\bra{z=\avg{\bm{\Theta}}^\top\mrm{Cov}[\bm{\Theta}]^{-1}\avg{\bm{\Theta}}\ge 0}\\
&\Leftrightarrow 1-2z/\Sigma\ge 0\\
&\Leftrightarrow \avg{\bm{\Theta}}^\top\mrm{Cov}[\bm{\Theta}]^{-1}\avg{\bm{\Theta}}\le\frac{\Sigma}{2}.
\end{aligned}
\end{equation}

\subsection{TUR for active matter systems}\label{app:TUR.act.mat}

The dynamics in Eq.~\eqref{eq:map.und.dyn} can be obtained from Eq.~\eqref{eq:und.Lan} by plugging
$m_i\leftarrow\tau,~G_{ij}(\br)\leftarrow-\delta_{ij}-\tau\mu\pp_{r_ir_j}^2\Phi(\br),~F_i(\br)\leftarrow-\mu\pp_{r_i}\Phi(\br)$.
According to Eq.~\eqref{eq:raw.TUR.2}, we have
\begin{equation}
\frac{\mrm{Var}[\Theta]}{\avg{\Theta}^2}\geq\frac{2}{\mca{T}(9\sigma+4\Upsilon_a)+\Omega},\label{eq:raw.TUR.3}
\end{equation}
where $\Omega$ is defined as in Eq.~\eqref{eq:omgs} and $\Upsilon_a$ is given by
\begin{equation}
\Upsilon_a=\sum_{i=1}^N\frac{1}{D_i}\brah{\avg{F_i(\br)^2}-2\avgl{\bram{\sum_{j=1}^Nv_jG_{ij}(\br)}^2}+\sum_{j=1}^N\avg{v_jF_i(\br)G_{ij}(\br)}-\frac{3D_i}{m_i}\avg{G_{ii}(\br)}}.\label{eq:ups2}
\end{equation}
In the steady state, we have
\begin{align}
0&=\sum_{i=1}^{N}\bras{\pp_{r_i}J^{\rm ss}_{r_i}(\xs)+\pp_{v_i}J^{\rm ss}_{v_i}(\xs)}\nonumber\\
&=\sum_{i=1}^N\bras{\pp_{r_i}\bras{v_i\pss(\xs)}+\frac{1}{m_i}\pp_{v_i}\bras{\sum_{j=1}^Nv_jG_{ij}(\br)\pss(\xs)}+\frac{1}{m_i}\pp_{v_i}\bras{F_i(\br)\pss(\xs)}-\frac{D_i}{m_i^2}\pp_{v_i}^2\pss(\xs)}.\label{eq:rel.ss.Fok}
\end{align}
Using the relation in Eq.~\eqref{eq:rel.ss.Fok} and taking integration by parts, we obtain for $i\neq j$
\begin{equation}
\avg{v_jF_i(\br)G_{ij}(\br)}=-m_i\sum_{k}\avgl{v_iv_k\pp_{r_k}F_i(\br)}-\sum_{l\neq j}\avgl{v_lF_i(\br)G_{il}(\br)}-\avg{F_i(\br)^2}.
\end{equation}
Hence,
\begin{equation}
\sum_{j=1}^N\avg{v_jF_i(\br)G_{ij}(\br)}=-\avg{F_i(\br)^2}-m_i\sum_{j=1}^N\avgl{v_iv_j\pp_{r_j}F_i(\br)}.\label{eq:aux.equ1}
\end{equation}
By plugging Eq.~\eqref{eq:aux.equ1} into Eq.~\eqref{eq:ups2}, we have
\begin{equation}
\Upsilon_a=-\sum_{i=1}^N\frac{1}{D_i}\brah{m_i\sum_{j=1}^N\avgl{v_iv_j\pp_{r_j}F_i(\br)}+2\avgl{\bram{\sum_{j=1}^Nv_jG_{ij}(\br)}^2}+\frac{3D_i}{m_i}\avg{G_{ii}(\br)}}.\label{eq:phi}
\end{equation}
Next, substituting $m_i\leftarrow\tau,~G_{ij}(\br)\leftarrow-\delta_{ij}-\tau\mu\pp_{r_ir_j}^2\Phi(\br),~F_i(\br)\leftarrow-\mu\pp_{r_i}\Phi(\br)$ into Eq.~\eqref{eq:phi}, we obtain
\begin{equation}
\Upsilon_a=\sum_{i=1}^N\frac{1}{D_i}\brah{\tau\mu\sum_{j=1}^N\avgl{v_iv_j\pp_{r_ir_j}^2\Phi(\br)}-2\avgl{\bram{\sum_{j=1}^Nv_j\bras{\delta_{ij}+\tau\mu\pp_{r_ir_j}^2\Phi(\br)}}^2}+\frac{3D_i}{\tau}\avg{1+\tau\mu\pp_{r_i}^2\Phi(\br)}}.
\end{equation}
The TUR for the active matter system is then given as follows:
\begin{equation}
\frac{\mrm{Var}[\Theta]}{\avg{\Theta}^2}\ge\frac{2}{\Sigma},
\end{equation}
where $\Sigma=\mca{T}(9\sigma+4\Upsilon_a)+\Omega$.

\end{widetext}


%

\end{document}